\documentclass[aps,pra,twocolumn]{revtex4}
\usepackage{amsmath}
\usepackage{amssymb}
\newcommand\Ho{{\hat H}}
\newcommand\Ko{{\hat K}}
\newcommand\Vo{{\hat V}}
\newcommand\ro{{\hat\rho}}
\newcommand\rb{\boldsymbol{r}}
\newcommand\Zo{\boldsymbol{\hat Z}}
\newcommand\xo{\boldsymbol{\hat x}}
\newcommand\po{\boldsymbol{\hat p}}
\newcommand\Jo{\boldsymbol{\hat J}}

\newcommand\no{{\hat n}}
\newcommand\tr{\mbox{\rm tr}}

\begin{document}
\title{On Conserved Current in Markovian Open Quantum Systems}

\author{Andr\'as Bodor}
\affiliation{
Department of the Physics of Complex Systems,\\
E\"otv\"os University; H-1117 Budapest, Hungary}

\author{Lajos Di\'osi}
\email{diosi@rmki.kfki.hu}
\homepage{www.rmki.kfki.hu/~diosi} 
\affiliation{Research Institute for Particle and Nuclear Physics\\
H-1525 Budapest 114, P.O.Box 49, Hungary}

\date{\today}

\begin{abstract}
We reexamine the markovian approximation of local current in open quantum systems, 
discussed recently by Gebauer and Car. Our derivation is more transparent, the proof 
of current conservation becomes explicit and easy.
\end{abstract}


\maketitle
\section{Introduction}
Quantum mechanics is reversible at the microscopic level. If, however, an atomic system is coupled 
to a reservoir the atomic reduced dynamics becomes irreversible and the corresponding evolution equation 
contains a memory term. Yet, in Markov approximation the time evolution of the atomic reduced density 
matrix $\ro$ satisfies the memory-less master equation:
\begin{equation}\label{Master}
\frac{d\ro}{dt}=-i[\Ho,\ro]+\mathcal{L}\ro,
\end{equation}
where $\Ho$ is the atomic Hamiltonian. The irreversible term can be written into the Lindblad-form
\cite{Lin76},\cite{GKS76}:
\begin{equation}\label{L}
\mathcal{L}\ro=\sum_i (\Vo_i^\dagger\ro\Vo_i-\Vo_i\Vo_i^\dagger\ro)g_i + H.C.,
\end{equation}
where $\{\Vo_i\}$ are certain atomic operators and $\{g_i\}$ are complex coefficients satisfying
$\mathrm{Re}~g_i>0$.
This equation constitutes the phenomenological dynamics of the given markovian open quantum system. 

In Hamiltonian dynamics with \emph{local} potentials and interactions
an important conserved quantity is the particle density. Consider, for simplicity's sake, 
a single atomic electron with the canonical pair of operators $\xo,\po$.
The electron's density operator at location $\rb$ reads:
\begin{equation}\label{n}
\no(\rb)=\delta(\rb-\xo)=\left|\rb\right>\left<\rb\right|,
\end{equation}
where the position eigenstates satisfy the equation $\xo\left|\rb\right>=\rb\left|\rb\right>$.
There is a local current defined by:
\begin{equation}\label{J}
\Jo(\rb)=\frac{1}{2m}[\po\no(\rb)+\no(\rb)\po].
\end{equation} 
The current $\Jo$ is conserved, i.e., the Heisenberg operators $\no_t,\Jo_t$ satisfy the continuity equation:
\begin{equation}\label{cont}
\frac{d\no_t(\rb)}{dt}+\boldsymbol{\nabla}\Jo_t(\rb)=0.
\end{equation}
A similar continuity equation follows for the expectation values as well. The continuity equation still
holds if the atom interacts with other systems provided the interaction is \emph{local}. Let us, indeed,
suppose a certain local interaction with a reservoir. The atom becomes an open system governed 
by the master equation (\ref{Master}) in the markovian approximation. 
Let us write down the continuity equation for the expectation values of $\no$ and $\Jo$ in the atomic state 
$\ro$, and substitute eq.(\ref{Master}). In Schr\"odinger picture we obtain the following:
\begin{equation}\label{contL}
\frac{d \langle\no(\rb)\rangle}{dt}+\boldsymbol{\nabla}\langle\Jo(\rb)\rangle
=\langle\mathcal{L}^\star\no(\rb)\rangle.
\end{equation}
We see that the extra term $\langle\mathcal{L}^\star\no(\rb)\rangle\equiv\tr[\no(\rb)\mathcal{L}\ro]$
on the r.h.s. may in general violate the local conservation of the current.
Gebauer and Car \cite{GebCar04} noticed that such violation is a consequence of Markov approximation.
They also pointed out that current conservation can be restored by adding a dissipative correction $\Jo_D$ to 
the Hamiltonian current $\Jo$. These authors have derived $\Jo_D$ from the exact reversible dynamics of
the atom+reservoir. Here we re-consider the issue and propose a shorter and transparent derivation. Our 
formulating $\Jo_D$ is fairly explicit to satisfy local conservation. This latter is the main progress with
respect to ref.\cite{GebCar04} which would need subtle and lengthier proof \cite{GebCar_priv}, not even published 
in ref.\cite{GebCar04} or elsewhere. 

It is clear from eq.(\ref{contL}) that $\Jo+\Jo_D$ will satisfy the continuity equation if:
\begin{equation}\label{JDconserv}
\boldsymbol{\nabla}\Jo_D(\rb)=\mathcal{L}^*\no(\rb).
\end{equation}
In one dimension with boundary conditions \mbox{$\Jo_D(\pm\infty)=0$}
the above equation has a unique solution 
\mbox{$\Jo_D(r)=\int_{-\infty}^r dr'\mathcal{L}^\star\no(r')$}
but in higher dimensions additional considerations are necessary
to uniquely determine $\Jo_D$. In fact, the Markov dynamics (\ref{Master}) of the atomic state $\ro$ is not 
sufficient to calculate $\Jo_D$. We must carefully inspect the Markov approximation of the
exact current. In Sec.II we review the derivation of
the master equation (\ref{Master}) in the usual Born-Markov approximation. Then Sec.III contains
our derivation of the dissipative current $\Jo_D$.  
 
\section{Born-Markov approximation}
Let stand $\Ho+\Ko+\Ho_R$ for the total Hamiltonian where
$\Ho$ belongs to the atomic system, $\Ho_R$ belongs to the reservoir.
The interaction Hamiltonian is local, we consider the simple case 
$\Ko=\Vo(\xo)\hat F$ where the ``field'' $\hat F$ is a reservoir operator \cite{Foot}. 
We suppose that the initial state of the whole system is 
$\ro\ro_R$ where $\ro_R$ is the reservoir equilibrium state.
We denote by $\hat O_\tau$ the operator $\hat O$ at time $\tau$ in interaction picture.
Let us introduce the reservoir correlation function
\begin{equation}\label{corr}
g(\sigma-\tau)=\tr_R\left(\hat F_\sigma\hat F_\tau\ro_R\right),
\end{equation}
which is a complex non-negative time-translation invariant kernel.
Consider the change of the atomic reduced density matrix during a certain time $\Delta$. 
In interaction picture, second order perturbation theory yields:
\begin{eqnarray}\label{pre-Lindblad}
\frac{\ro_\Delta-\ro}{\Delta}=
\frac{1}{\Delta}\int_{0}^{\Delta}\!\!\!\!\!\!d\sigma \int_{0}^{\sigma}\!\!\!\!\!d\tau
\left(\Vo_\tau\ro\Vo_\sigma - \Vo_\sigma\Vo_\tau\ro\right)g(\sigma-\tau)\nonumber\\+H.C.
\end{eqnarray}
where the first order term in $\Vo$ vanishes since we assume $\tr_R(\hat F\ro_R)=0$. 
Suppose that  $\tau_A\gg\tau_R$ where $\tau_A$ is the time scale of atomic state 
evolution and $\tau_R$ is the reservoir correlation time. Choose $\Delta$ in between
the reservoir and atomic time-scales:
\begin{equation}\label{coarse}
\tau_R~~\ll~~\Delta~~\ll~~\tau_A.
\end{equation}
A standard way to proceed is the Born-Markov approximation that requires weak coupling and results in 
the master equation (\ref{Master}) for the atomic state $\ro_t$ 
\emph{time-coarse-grained on scale} $\Delta$. 

The implementation of the Born-Markov approximation has two steps. We identify the l.h.s.
of (\ref{pre-Lindblad}) by the time derivative $d\ro/dt$ of the time-coarse-grained
atomic state $\ro$. On the r.h.s., we extend the integration limit as well as the denominator $\Delta$ 
to $\infty$. 
The latter limit is rigorously justified by a particular rescaling \cite{resc} where $\Vo\rightarrow0$ 
while $\Vo^2\Delta$ is constant \cite{DavGFVKS}. Accordingly, the explicit form of the Lindblad-superoperator 
becomes:
\begin{eqnarray}\label{Lindblad}
\mathcal{L}\ro=
\lim_{t\rightarrow\infty}
\frac{1}{t}\int_{0}^t\!\!\!\!\!d\sigma \int_{0}^{\sigma}\!\!\!\!\!\!d\tau
\left(\Vo_\tau\ro\Vo_\sigma - \Vo_\sigma\Vo_\tau\ro\right)g(\sigma-\tau)\nonumber\\+H.C.
\end{eqnarray}
It is easy to inspect that the r.h.s. is already of the Lindblad form (\ref{L}). To achieve the familiar
spectral representation, we expand the interaction potentials as
\begin{equation}
\Vo_\tau=\sum_\omega e^{i\omega\tau}\Vo_\omega\,,
\end{equation}
where the sum is over all possible transition frequencies $\omega=\epsilon_n-\epsilon_m$ of the
system Hamiltonian $\Ho=\sum_n\epsilon_n\vert n\rangle\langle n\vert$. Obviously, the spectral component 
of $\Vo$ is of this form:
\begin{equation}\label{Vspect}
\Vo_\omega=\Vo_{-\omega}^\dagger=\sum_{n,m\atop\epsilon_n-\epsilon_m=\omega} \vert n\rangle V_{nm}\langle m\vert.
\end{equation}
As $t\rightarrow\infty$ the oscillating terms in the integral (\ref{Lindblad}) drop out and we get
the master equation (\ref{Master}) where:
\begin{equation}
\mathcal{L}\ro=\sum_\omega (\Vo_{\omega}^\dagger\ro\Vo_{\omega}-\Vo_{\omega}\Vo_{\omega}^\dagger\ro)g_\omega^{(+)} + H.C.
\end{equation}
and
\begin{equation}
g_\omega^{(+)}= \int_{0}^{\infty} d\tau e^{i\omega\tau}g(\tau).
\end{equation}

\section{The markovian conserved current}
Constructing the correct markovian coarse-grained current $\Jo+\Jo_D$ needs special care.
First, we calculate the average of the exact current over the period $\Delta$ of time-coarse-graining.
We expect that it becomes the sum of the unperturbed contribution $\Jo$ and the
markovian correction $\Jo_D$ in the Born-Markov limit:
\begin{equation}\label{avJ1} 
\frac{1}{\Delta}\int_0^\Delta d\sigma\langle\Jo\rangle_\sigma\longrightarrow
\langle\Jo\rangle+\langle\Jo_D\rangle
\end{equation}
where $\langle\dots\rangle_\sigma$ stands for expectation value in the exact time-dependent
state evolved from $\ro\ro_R$ by the total Hamiltonian $\Ho+\Ho_R+\Ko$ while 
$\langle\dots\rangle$ stands for the expectation value in the coarse-grained state $\ro$. 
Let us introduce the time-dependent auxiliary variable $\Zo(\rb;t)$ defined as 
\begin{equation}\label{Zo} 
\Zo(\rb;t)=\int_0^t d\sigma\Jo_\sigma(\rb).
\end{equation}
Then, in the interaction picture, the l.h.s. of (\ref{avJ1}) can be written as:
\begin{equation}\label{avJ2} 
\frac{1}{\Delta}\tr{\Zo(\Delta)\ro_\Delta}
-\frac{1}{\Delta}\tr\int_0^\Delta d\sigma\Zo(\sigma)\frac{d\ro_\sigma}{d\sigma}. 
\end{equation}
\emph{In Born-Markov limit}, the time-coarse-grained atomic state $\ro_t$ evolves smoothly and only 
``differentially'' during time $\Delta$, so does the current $\Jo_\sigma$, too,  in such states:
we can thus write the first term as just $\langle\Jo\rangle$. In the second term,
\emph{before taking the Born-Markov limit}, we substitute the perturbative expression of $d\ro_\sigma/d\sigma$:
\begin{eqnarray}\label{avJ3} 
&~&-\frac{1}{\Delta}\tr\int_0^\Delta\!\!\!\!d\sigma\Zo(\sigma)\int_0^\sigma\!\!d\tau
\left(\Vo_\tau\ro\Vo_\sigma - \Vo_\sigma\Vo_\tau\ro\right)g(\sigma-\tau)\nonumber\\
&~&+H.C. 
\end{eqnarray}

Now we take the Born-Markov limit which, again, allows us to calculate the r.h.s. by taking the limit
$\Delta\rightarrow\infty$:
\begin{eqnarray}\label{avJ4} 
&~&\langle\Jo_D\rangle=\nonumber\\
&-&\lim_{t\rightarrow\infty}
\frac{1}{t}\tr\int_{0}^t\!\!\!\!\!d\sigma \int_{0}^{\sigma}\!\!\!\!\!\!d\tau
\Zo(\sigma)
\left(\Vo_\tau\ro\Vo_\sigma - \Vo_\sigma\Vo_\tau\ro\right)g(\sigma-\tau)\nonumber\\
&+&H.C. 
\end{eqnarray}
We can express the dissipative current itself:
\begin{eqnarray}\label{JD}
&~&\Jo_D=\nonumber\\
&-&\lim_{t\rightarrow\infty}
\frac{1}{t}\int_{0}^t\!\!\!\!\!d\sigma \int_{0}^{\sigma}\!\!\!\!\!\!d\tau
\left(\Vo_\sigma\Zo(\sigma)\Vo_\tau - \Zo(\sigma)\Vo_\sigma\Vo_\tau\right)g(\sigma-\tau)\nonumber\\
&+&H.C. 
\end{eqnarray}
Recall that $\Zo(\rb;\sigma)=\int_0^\sigma d\lambda\Jo_\lambda(\rb)$ so that $\Jo_D(\rb)$ has turned
out to be a local linear functional of the unperturbed local current $\Jo(\rb)$.

It is easy to prove that $\Jo_D$ contributes properly to current conservation. Let us
take the divergence of both sides of eq.(\ref{JD}). Observe that 
\begin{equation}
\nabla\Zo(\sigma)=\no_\sigma-\no,
\end{equation}
because density conservation $\boldsymbol{\nabla}\Jo_\sigma+d\no_\sigma/d\sigma=0$ holds for the 
unperturbed atomic current and density. 
We have assumed that the interaction is local, so $[\no_\sigma,\Vo_\sigma]=0$.
Therefore the terms proportional to $\no_\sigma$ cancel each other
on the r.h.s. of eq.(\ref{JD}) and we obtain: 
\begin{eqnarray}\label{divJD}
&~&\boldsymbol{\nabla}\Jo_D=\\
&~&\lim_{t\rightarrow\infty}
\frac{1}{t}\int_{0}^t\!\!\!\!\!d\sigma \int_{0}^{\sigma}\!\!\!\!\!\!d\tau
\left(\Vo_\sigma\no\Vo_\tau - \no\Vo_\sigma\Vo_\tau\right)g(\sigma-\tau)~+H.C.\nonumber
\end{eqnarray}
The r.h.s. coincides with $\mathcal{L^\star}\no$ as it should, cf. eq.(\ref{JDconserv}).

\section{Closing remarks}
Our central result is the expression (\ref{JD}) of the dissipative current $\Jo_D$.
Its form was convenient when we proved the continuity equation from the locality of the interaction.
In concrete calculations, however, the spectral representation is the preferred one. 
As the spectral representation of $\Zo(\sigma)$ is 
\begin{equation}
\Zo(\sigma)=\sum_{\omega_J}\Jo_{\omega_J}\frac{e^{i\omega_J\sigma}-1}{i\omega_J}\,,
\end{equation}
eq.(\ref{JD}) yields two terms for the spectral expression of $\Jo_D$.
To compare it with the result of Car and
Gebauer \cite{GebCar04}, we present the spectral expression of the expectation value:
\begin{eqnarray}\label{JDomega} 
\langle\Jo_D\rangle&=&
\tr \left(\sum_{\substack{\omega_J+\omega_1-\omega_2=0\\ \omega_\rho=0}}-
      \sum_{\substack{\omega_1-\omega_2=0\\ \omega_J+\omega_\rho=0}}
\right)\\
&~&i\frac{\Jo_{\omega_J}}{\omega_J}\left(\Vo_{\omega_2}^\dagger\ro_{\omega_\rho}\Vo_{\omega_1}-
\Vo_{\omega_1}\Vo_{\omega_2}^\dagger\ro_{\omega_\rho}\right)g_{\omega_2}^{(+)}~+~H.C.\nonumber 
\end{eqnarray}
Here $\Jo_{\omega_J},\ro_{\omega_\rho}$ and $\Vo_\omega$ are the respective spectral representations of $\Jo,\ro$ and
$\Vo$, cf. eq.(\ref{Vspect}). The two sums cancel the singularity at $\omega_J=0$.
We note without going into the details that, for non-degenerate transition frequencies, 
our result (\ref{JDomega}) coincides with eq.(16) in \cite{GebCar04}
apart from minor typos of the latter \cite{GebCarEquiv}. 

Our result allows us to calculate the markovian correction $\Jo_D$ to the local current in open
quantum systems with discrete spectrum. The method applies to quantum dots directly \cite{qdots}. 
The case of quantum brownian motion, however, requires a 
suitably modified approach to cope with the markovian limit of an open system with continuous spectrum.

L. D. work was supported by the Hungarian OTKA Grant No. 49384.
The authors would like to thank Tam\'as Geszti for discussions and encouragement.
We appreciate a helpful earlier correspondence with Ralph Gebauer and Roberto Car.


\begin{thebibliography}{99}
\bibitem{Foot} The generic form of $\Ko$ is \mbox{$\Vo_1(\xo)\hat F_1+ \Vo_2(\xo)\hat F_2+\dots$} and our results
can immediately be extended to such general interactions.
\bibitem{resc} Temporarily, re-write the interaction Hamiltonian into the form $\lambda\Ko$ where $\lambda$ is
the parameter of coupling. In the limit $\lambda\rightarrow0$, the second-order perturbative expression yields 
the exact time-derivative of the coarse-grained atomic state $\ro$ for times of order $\lambda^{-2}$ \cite{DavGFVKS}. 
Accordingly, the r.h.s. of eq.(\ref{pre-Lindblad}) can equivalently be calculated if make the subsitutions
$\Ko\rightarrow\lambda\Ko$ and $\Delta\rightarrow\lambda^{-2}\Delta$ at $\lambda\rightarrow0$. This is
equivalent with the limit indicated on the r.h.s. of eq.(\ref{Lindblad}).
\bibitem{GebCarEquiv}
For the reader's convenience, let us match the
structure and notations of our eq.(\ref{JDomega}) with those of \cite{GebCar04}.
The first and the second sums of the r.h.s. are related to $\mathcal{T}_{kl}$ and $\mathcal{R}_{kl}$ in
\cite{GebCar04}. In our notations:
\begin{equation}
\mathcal{T}_{kl}=\sum_{\substack{\omega_1-\omega_2=\epsilon_k-\epsilon_l\\ \omega_\rho=0}}
\left(\Vo_{\omega_2}^\dagger\ro_{\omega_\rho}\Vo_{\omega_1}-
\Vo_{\omega_1}\Vo_{\omega_2}^\dagger\ro_{\omega_\rho}\right)g^{(+)}_{\omega_2} + H.C.
\nonumber
\end{equation}
In the correct version of \cite{GebCar04}:
\begin{equation}
\mathcal{T}_{kl}=\sum_{n\neq k}\left(\Vo_{kn}\bar{S}_{nn}(t)\Vo_{nl}w^+_{nk}-
\Vo_{kn}\Vo_{nl}\bar{S}_{ll}(t)w^+_{ln}\right) + H.C.
\nonumber
\end{equation}
Concerning $\mathcal{R}_{kl}$ we note that if there are no transition frequency degeneracies
then $\Vo_{\omega_1}^\dagger\ro_{\omega_\rho}\Vo_{\omega_1}$ has nonvanishing terms
only if $\omega_\rho=0$. So the sandwich term drops out and  that is why $\mathcal{R}_{kl}$ in
\cite{GebCar04} has only two terms.
\end{thebibliography}
\end{document}